# ANTITRUST, INTELLECTUAL PROPERTY AND STANDARD SETTING ORGANIZATIONS[1]

**Mark A. Lemley**[2]
UNIVERSITY OF CALIFORNIA AT BERKELEY

***Author's Note**: This is a partial version of a much longer work in progress. Because of the space limits imposed by the conference, I have chosen to submit only the introduction and Parts I and II of the paper. The abstract spells out the agenda for the rest of the paper. Part II offers an empirical analysis of the intellectual property policies of standard-setting organizations in the telecommunications and computer networking industries. My presentation will discuss the full paper, and I will make it available upon request.*

***Abstract:** The role of institutions in mediating the use of intellectual property rights has long been neglected in debates over the economics of intellectual property. In a path-breaking work, Rob Merges studied what he calls "collective rights organizations," industry groups that collect intellectual property rights from owners and license them as a package. Merges finds that these organizations ease some of the tensions created by strong intellectual property rights by allowing industries to bargain from a property rule into a liability rule. Collective rights organizations thus play a valuable role in facilitating transactions in intellectual property rights.*

*There is another sort of organization that mediates between intellectual property owners and users, however. Standard-setting organizations (SSOs) regularly encounter situations in which one or more companies claim to own proprietary rights that cover a proposed industry standard. The industry cannot adopt the standard without the permission of the intellectual property owner (or owners).*

*How SSOs respond to those who assert intellectual property rights is critically important. Whether or not private companies retain intellectual property rights in group standards will determine whether a standard is "open" or "closed." It will determine who can sell compliant products, and it may well influence whether the standard adopted in the market is one chosen by a group or one offered by a single company. SSO rules governing intellectual property rights will also affect how standards change as technology improves. To give just one example, the Internet runs on a set of open, non-proprietary protocols in large part because the Internet Engineering Task Force (IETF), the SSO that controls the TCP and IP protocols, had a long-standing policy that it would not adopt proprietary standards. That policy has now changed; it remains to be seen whether the open nature of the Internet will change with it. But in any event, the magnitude of the stakes should be clear.*

*Given the importance of SSO rules governing intellectual property rights, there has been surprisingly little treatment of SSOs or their intellectual property rules in the legal literature. My aim in this article is to fill that void. To do so, I have surveyed*

---

[2] Professor of Law, University of California at Berkeley (Boalt Hall); of counsel, Fish & Richardson PC. Thanks to Ryan Garcia for research assistance, and to the Open Group, Sun Microsystems, Erv Basinski and Carl Cargill for helping me navigate the thicket of standard-setting organizations in computer networking.

*the intellectual property policies of dozens of SSOs, primarily but not exclusively in the computer networking and telecommunications industries. In Part I, I provide some background on SSOs themselves, and discuss the value of group standard setting in network markets. In Part II, I discuss my empirical research, which demonstrates a remarkable diversity among SSOs in how they treat intellectual property. In Part III, I consider the constraints the antitrust laws place on SSOs in general, and on their adoption of intellectual property policies in particular. In Part IV, I analyze a host of unresolved legal issues relating to the applicability and enforcement of such intellectual property policies. Finally, in Part V I offer recommendations to scholars, policy-makers, companies, and SSOs for dealing with SSO intellectual property policies.*

*In the end, I hope to convince the reader of four things. First, SSO rules governing intellectual property fundamentally change the way in which we must approach the study of intellectual property. It is not enough to consider IP rights in a vacuum; we must consider them as they are actually used in practice. And that means considering how SSO rules affect IP incentives. Second, there is a remarkable diversity among SSOs in how they treat IP rights. This diversity is largely accidental, and does not reflect conscious competition between different policies. Third, antitrust law is not well designed to take account of the modern role of SSOs. Antitrust rules may unduly restrict SSOs even when those organizations are serving procompetitive ends. Fourth, the enforcement of SSO IP rules presents a number of important but unresolved problems of contract and intellectual property law. I offer some suggestions for how to think about these problems in the final part of this paper.*

The standard economic theory of intellectual property is well known. Intellectual creations are public goods, much easier and cheaper to copy than they are to produce in the first place. Absent some form of exclusive right over inventions, no one (or not enough people) will bother to innovate. Intellectual property rights are thus a "solution" to the public goods problem because they privatize the public good, and therefore give potential inventors an incentive to engage in research and development.[3]

In the real world, things aren't so simple. People innovate for lots of reasons, and the existence of intellectual property rights doesn't appear to be chief among them.[4] Intellectual property rights have different impact on different industries, depending on the nature and cost of innovation, the maturity of the industry, and the relationship between patentable inventions and marketable products. These different characteristics, coupled with uncertainty about how much incentive intellectual property rights actually give, lead to vigorous debates about the wisdom of intellectual property rights in particular contexts, notably software and electronic commerce.[5]

The role of institutions in mediating the use of intellectual property rights has long been neglected in these debates. In a path-breaking work, Rob Merges studied what he calls "collective rights organizations," industry groups that collect intellectual property rights from owners and license them as a package.[6] Merges finds that these organizations ease some of the tensions created by strong intellectual property rights by allowing industries to bargain from a property rule into a liability rule.[7] Collective rights organizations thus play a valuable role in facilitating transactions in intellectual property rights.

---

[3] For a discussion of the standard theory, see Mark A. Lemley, *The Economics of Improvement in Intellectual Property Law*, 75 **Tex. L. Rev.** 989 (1997).

[4] *See, e.g.,* Richard C. Levin et al, *Appropriating the Returns from Industrial Research and Development*, 1987 BROOKINGS PAPERS ON ECON. ACTIVITY 783; Wesley M. Cohen, *et al.*, *Appropriability Conditions and Why Firms Patent and Why They Do Not in the American Manufacturing Sector*, presented at the Stanford Workshop on Intellectual Property and Industry Competitive Standards, Stanford Law School, April 17-18, 1998.

[5] *See, e.g.,* Julie E. Cohen & Mark A. Lemley, *Patent Scope and Innovation in the Software Industry*, 89 **Calif. L. Rev.** 1 (2001); Pamela Samuelson et al., *A Manifesto Concerning the Legal Protection of Computer Programs*, 94 **Colum. L. Rev.** 2431 (1994).

[6] Robert P. Merges, *Institutions Supporting Transactions in Intellectual Property Rights*, 84 **Calif. L. Rev.** 1293 (1996). Merges discusses two basic sorts of collective rights organizations: patent pools and music licensing collectives.

There is another sort of organization that mediates between intellectual property owners and users, however. Standard-setting organizations (SSOs) regularly encounter situations in which one or more companies claim to own proprietary rights that cover a proposed industry standard. The industry cannot adopt the standard without the permission of the intellectual property owner (or owners).

How SSOs respond to those who assert intellectual property rights is critically important. Whether or not private companies retain intellectual property rights in group standards will determine whether a standard is "open" or "closed." It will determine who can sell compliant products, and it may well influence whether the standard adopted in the market is one chosen by a group or one offered by a single company. SSO rules governing intellectual property rights will also affect how standards change as technology improves. To give just one example, the Internet runs on a set of open, non-proprietary protocols in large part because the Internet Engineering Task Force (IETF), the SSO that controls the TCP and IP protocols, had a long-standing policy that it would not adopt proprietary standards. That policy has now changed; it remains to be seen whether the open nature of the Internet will change with it. But in any event, the magnitude of the stakes should be clear.

Given the importance of SSO rules governing intellectual property rights, there has been surprisingly little treatment of SSOs or their intellectual property rules in the legal literature.[8] My aim in this article is to fill that void. To do so, I have surveyed the intellectual property policies of dozens of SSOs, primarily but not exclusively in the computer networking and telecommunications industries. In Part I, I provide some background on SSOs themselves, and discuss the value of group standard setting in network markets. In Part II, I discuss my empirical research, which demonstrates a remarkable diversity among SSOs in how they treat intellectual property. In Part III, I consider the constraints the antitrust laws place on SSOs in general, and on their adoption of intellectual property policies in particular. In Part IV, I analyze a host of unresolved legal issues relating to the applicability and enforcement of such intellectual property policies. Finally, in Part V I offer recommendations to scholars, policy-makers, companies, and SSOs for dealing with SSO intellectual property policies.

In the end, I hope to convince the reader of four things. First, SSO rules governing intellectual property fundamentally change the way in which we must approach the study of intellectual property. It is not enough to consider IP rights in a vacuum; we must consider them as they are actually used in practice. And that means considering how SSO rules affect IP incentives. Second, there is a remarkable diversity among SSOs in how they treat IP rights. This diversity is largely accidental, and does not reflect conscious competition between different policies. Third, antitrust law is not well designed to take account of the modern role of SSOs. Antitrust rules may unduly restrict SSOs even when those organizations are serving procompetitive ends. Fourth, the enforcement of SSO IP rules presents a number of important but unresolved problems of contract and intellectual property law. I offer some suggestions for how to think about these problems in the final part of this paper.

---

[7] *See, e.g.,* Guido Calabresi & A. Douglas Melamed, *Property Rules, Liability Rules, and Inalienability: One View of the Cathedral*, 85 **Harv. L. Rev.** 1089 (1972) (discussing the difference between property rule regimes, in which the owner of a right is entitled to an injunction, and liability rule regimes, in which a right is enforced only by paying damages to compensate for the owner's loss).

[8] The literature on antitrust and standard-setting organizations is voluminous, though much of it considers issues unrelated to intellectual property. Among the better sources are James J. Anton & Dennis A. Yao, *Standard-Setting Consortia, Antitrust, and High-Technology Industries*, 64 **Antitrust L.J.** 247, 248, 262-63 (1995); Mark A. Lemley, *Antitrust and the Internet Standardization Problem*, 28 **Conn. L. Rev.** 1041 (1996); H.S. Gerla, *Federal Antitrust Law and Trade and Professional Association Standards and Certification*, 19 **U. Dayton L. Rev.** 471 (1994); David Teece, *Information Sharing, Innovation, and Antitrust*, 62 **Antitrust L.J.** 465 (1994); Jack E. Brown, *Technology Joint Ventures to Set Standards or Define Interfaces*, 61 **Antitrust L.J.** 921 (1993); Hol & Badger, *The Antitrust Challenge to Non-Profit Certification Organizations: Conflicts of Interest and a Practical Rule of Reason Approach to Certification Programs as Industry-Wide Builders of Competition and Efficiency*, 60 **Wash. U.L.Q.** 357 (1982) (endorsing fact-specific rule of reason approach); Thomas A. Priaino, Jr., *The Antitrust Analysis of Network Joint Ventures*, 47 **Hastings L.J.** 5 (1995); Thomas M. Jorde & David J. Teece, *Rule of Reason Analysis of Horizontal Arrangements: Agreements Designed to Advance Innovation and Commercialize Technology*, 61 **Antitrust L.J.** 579 (1993); Mark Shurmer & Gary Lea, *Telecommunications Standardization and Intellectual Property Rights: A Fundamental Dilemma?,* in **Standards Policy for Information Infrastructure** 378 (Kahin & Abbate eds. 1995). The best work to date on intellectual property policies of standard setting organizations is Michael J. Schallop, *The IPR Paradox: Leveraging Intellectual Property Rights to Encourage Interoperability in the Network Computing Age*, 28 **AIPLA Q.J.** 195 (2000).

# I. The Nature and Importance of Standard Setting Organizations

*A. The Value of Standardization*

Standards (and standard-setting organizations) come in a variety of forms. I define a standard rather broadly, as any set of technical specifications which either does or is intended to provide a common design for a product or process. Some standards are extremely complex and technical in nature. For example, the set of applications programming interfaces that defines compatibility with the Microsoft Windows operating system is an industry standard; those who know and use the proper interfaces are compliant with the standard, and their products will "interoperate" with the Microsoft OS. But standards do not have to be so sophisticated. Ordinary consumers use a wide variety of standardized products in everyday life. In the U.S., electrical plugs and outlets are built to a particular standard for voltage, impedance, and plug shape. Without this standardization, no one could stay in a hotel room and have any confidence that their hair dryer would work in the hotel's outlet. The modern economy has also standardized telephone service, computer modem communication protocols, automobile ignition and transmission systems, and countless other products.

As these examples attest, in many markets standardization has significant consumer benefits. This is especially true in so-called "network markets," where the value of a product to a particular consumer is a function of how many other consumers use the same (or a compatible) product.[9] The paradigm example is the telephone network, in which the value of the product is entirely driven by the number of other people on the same network. Still other products -- like computer operating systems -- have some intrinsic value regardless of how many people use them, but gain value as more and more consumers adopt them. In these industries, consumers benefit from standardization not only because they can reliably use their product in a remote location, but also because they can exchange information with others who use the same standard. Further, markets for complementary products will often gear their production to work with a product that is an industry standard, rather than a product that has only a small market share. For example, software vendors are more likely to write computer applications programs compatible with Microsoft's operating system than with other operating systems, because there are more consumers for such a product. This in turn reinforces the desire of consumers to buy the product everyone else buys, a phenomenon known as "tipping."[10]

In network markets, then, standardization may well be inevitable, and certainly carries substantial consumer benefits. Even in non-network markets, standard-setting can have a variety of procompetitive and other beneficial effects. Agreeing on a set of standards can facilitate a competitive market for replacement parts or service in durable goods industries, for example. Further, in many industries standards may be valuable for reasons unrelated to or even inimical to competition. Construction products must meet industry standards for fire resistance, for example, and doctors, lawyers and many other professionals must meet minimum licensing standards. These latter standards are not procompetitive in the narrow sense of encouraging price competition; indeed, they may have the opposite effect. But standards of this type can still promote social welfare by ensuring that imperfect information does not lead consumers to buy dangerous products or hire unqualified doctors simply because they cost less.[11]

---

9 For literature on network effects, see, e.g., Mark A. Lemley & David McGowan, *Legal Implications of Network Economic Effects*, 86 **Calif. L. Rev.** 479 (1998); Michael Katz & Carl Shapiro, *Network Externalities, Competition, and Compatibility*, 75 **Am. Econ. Rev.** 424 (1985); Joseph Farrell & Garth Saloner, *Standardization, Compatibility, and Innovation*, 16 **Rand J. Econ.** 70 (1985); S.J. Liebowitz & Stephen E. Margolis, *Network Externality: An Uncommon Tragedy*, 8 **J. Econ. Persp.** 133 (1994).

10 *See* Katz & Shapiro, *supra* note [9].

11 Whether this sort of justification renders an otherwise anticompetitive agreement legal is a matter of some debate. On the one hand, the Supreme Court seemed to rule out any antitrust defense based along these lines in *National Society of Professional Engineers v. United States*, 435 U.S. 679 (1978) ("the Rule of Reason does not support a defense based on the assumption that competition itself is unreasonable."). On the other hand, many lower courts have recognized such a defense, holding at least that it precludes per se illegality. *See, e.g., Wilk v. American Medical Ass'n*, 719 F.2d 207, 221 (7th Cir. 1982) ("patient care" defense raised by organization required rule of reason treatment); *Kreuzer v. American Academy of Periodontology*, 735 F.2d 1479, 1493-94 (D.C. Cir. 1984) (rule restricting practice outside stated medical specialty subject to rule of reason analysis).

Because it does not directly concern intellectual property cases, resolution of this debate is outside the scope of this article.

On the other hand, standardization also poses some potential threats to competition. Absent network effects, economists generally presume that consumers fare best when many companies compete to offer different sorts of products. To the extent that standardization on a single product reduces consumer choice, it may be undesirable. Of course, if a market is truly competitive, unnecessary standardization should eventually be competed away by new entrants offering different sorts of products. But standard-setting organizations may be able to impede such competition, in effect acting as a cartel with the power to exclude output.[12] The general nature of this problem is discussed in more detail below.[13]

*B. The Benefits of Group Standard-Setting*

It remains, however, to consider the organizational form standardization may take. One approach to achieving interoperable standards is for a private industry organization open to all members to adopt a single standard. If the members of such a group collectively have a significant market share, their adoption of a standard may produce the "tipping" effect described above, bringing the rest of the industry into line.[14]

Not all standards are created by private standard-setting organizations, however. Two other organizational forms are worth considering. First, a standard may arise from the operation of the market, as consumers gravitate towards a single product or protocol and reject its competitors. This form of "de facto" standardization is particularly likely in markets characterized by strong network effects, because of the large benefits associated with adopting the same product everyone else does. To take just one example, the Microsoft operating systems are clearly de facto standards. No standard-setting organization "adopted" them as the preferred or official operating systems, but the market clearly chose Microsoft as the winner of a standards competition.

Another possibility is that the government might identify and set the appropriate standards and compel all participants in the market to comply. The government does this from time to time. For example, the Federal Communications Commission sets standards for interconnection between telephone networks and standards governing the use of products that might interfere with broadcast communications.[15] In the 1990s, the United States government stepped into the debate over the proper standard for high definition television (HDTV), selecting a standard that unified U.S. development work but was at odds with other standards adopted in Japan and Europe.[16] And government agencies such as the Advanced Research Projects Agency and the National Science Foundation played a crucial role in the development of the Internet, including the creation of Internet interconnection protocols. Indeed, some private Internet standard-setting groups such as InterNIC and the IETF were once government-sponsored standards organizations.

In this article I shall primarily be concerned with the activities of private standard-setting organizations. While de facto standards do raise significant antitrust issues relating to intellectual property, they are analytically distinct from the ones I discuss here.[17] Generally speaking, a de facto standard will be proprietary unless the standard-setter chooses to release it to the public. Government-set standards also present a very different set of issues, in part because of the state action and petitioning immunity doctrines.[18]

*C. Relationship to Intellectual Property*

Briefly stated, the issues in this article arise when a standard-setting organization adopts (or fails to adopt) a standard that is covered in whole or in part by an intellectual property right, generally but not necessarily an intellectual property right owned by a party that has some dealings with the organization.[19] These issues arise in three basic ways. First, a party upset with the decision to adopt (or not to adopt) a standard may challenge that decision as anticompetitive. In the intellectual property context, these claims are either by third parties against a standard-setting organization that has adopted a proprietary standard, or by patentees upset that a standard-setting organization has *not* adopted their proprietary standard. Second, a defendant may be charged with manipulating the standard-setting process, normally in order to gain adoption of a proprietary technology but conceivably in order to prevent its adoption. Finally, standard-setting organizations frequently use formal or informal mechanisms such as rules governing the ownership of intellectual property or joint defense arrangements to lessen the control an intellectual property owner has over a standard they adopt. These arrangements may themselves be challenged as anticompetitive. Because this last sort of agreement is of particular importance to intellectual property as well as antitrust policy, I discuss these policies in much more detail in Part II.

---

[12] Sometimes this power is economic, and results from the participation in the standard-setting organization of the largest companies in the industry. But some standard-setting organizations may wield direct legal control over a market, either directly (as where the courts delegate to bar associations the power to control entry into the profession) or indirectly (where a private standard-setting organization adopts standards that are routinely enacted into law by legislatures or city councils).

[13] *See infra* notes __-__ and accompanying text. For more detail, see XIII Hovenkamp, **Antitrust** at ¶ 2231b.

[14] Of course, not all standard-setting groups have such market control. As Libicki observes, many of the most successful group standards started small and grew to become dominant. *See* Martin C. Libicki, *Standards: The Rough Road to the Common Byte, in* Standards Policy for Information Infrastructure 35, 75 (Kahin & Abbate eds. 1995); *see also* Jim Isaak, *Information Infrastructure Meta-Architecture and Cross-Industry Standardization*, *in* Standards Policy for Information Infrastructure 100, 101 (arguing that group or open standards "must also reach the status of being 'de facto' to be sufficient").

[15] *See* F.C.C. Rules, 47 C.F.R. § 68.1.

[16] *See* Denise Caruso, *Debate Over Advanced TV Gives the F.C.C. a Chance to Be Assertive*, N.Y. Times, June 17, 1996, at D5; *F.C.C. Proposes Standards for Digital Television*, N.Y. Times, May 10, 1996, at D4.

[17] For a discussion, see Mark A. Lemley, *Antitrust and the Internet Standardization Problem*, 28 **Conn. L. Rev.** 1041 (1996).

[18] For more detail on these doctrines, see I Areeda & Hovenkamp at ¶200-231.

## II. How Standard Setting Organizations Treat Intellectual Property Rights

*A. Organizations Studied*

To see how standard-setting organizations treat intellectual property rights, I surveyed the rules and by-laws of 29 different standard-setting organizations. The organizations I chose were ones likely to be encountered by companies in the telecommunications and computer networking industries, where many of the most contentious intellectual property issues arise. They include both large national or international groups such as ANSI (the American National Standards Institute) and the ISO (the International Organization for Standardization), as well as smaller groups centered within the industries themselves. However, the collection of organizations here is by no means comprehensive, even within the telecommunications and computer networking industries.

I sought to identify several pieces of information with respect to an organization's policy on intellectual property rights. The first question was whether the organization had any policy at all regarding intellectual property. If they did, I then sought to determine whether the policy covered only patents, or whether it covered other forms of IP rights as well.

For those organizations that had policies governing IP, I sought to categorize the policy according to several factors. First, I determined whether the policy required disclosure of an IP right (as well as certain subsidiary questions, such as the nature of the search obligation (if any) and whether disclosure extended to pending as well as issued patents). Second, I determined the effect of an IP disclosure on the standard-setting process under the policy: chiefly whether the organization would refuse to adopt a standard covered by a patent. Finally, I determined whether the organization imposed a licensing requirement on intellectual property owners, and if so the nature of that requirement.

*B. Results*

1. Summary of Organization Policies

The full data from this survey are reprinted in Table 1 at the end of this article.

What is most striking about this data is the significant variation in policies among the different organizations. Of the 29 organizations I studied, 21 have written policies governing the ownership of intellectual property rights, 7 have no policy,[20] and one has a policy that was still in development. Most groups without a policy are small, industry-specific groups; all of the large standard-setting organizations I studied have well-developed policies in this area.

The subject matter of those policies varies significantly from group to group. Virtually all groups have either an express or implied obligation that members disclose intellectual property rights of which they were aware. Those groups that do not require disclosure generally impose other conditions that obviated the need for disclosure. Most commonly, they require royalty-free licensing of *all* member intellectual property rights that cover a group standard, whether or not it was disclosed to the organization.[21]

---

[19] If a standard-setting organization adopts as a standard a technical design covered by a patent owned by a non-member, only a more limited set of antitrust issues arise. The intellectual property owner is entitled to enforce its patent against those who use the standard. By contrast, refusal to adopt a standard covered by a patent owned by a third party could present antitrust issues, which I discuss *infra* notes __-__ and accompanying text.

[20] At least, no policy that either I or my research assistant could find. Some organizations may make their policy available only to members.

There is greater variation, however, with respect to *what* must be disclosed. While virtually all the policies I studied cover patents, a significant minority also cover copyright and trademark rights, or refer globally to "intellectual property rights" subject to the policy. Where patents are concerned, most organizations consider only issued patents. There is rarely discussion of the problem of pending patent applications.[22] A few organizations consider the issue, but do not require the disclosure of pending applications, which are ordinarily kept confidential. Two organizations, the ITU and OSGi, require disclosure of all pending patent applications. Two other organizations have an intermediate policy. The ATM Forum requires disclosure of published patent applications, but not unpublished ones.[23] And the WAP Forum requires disclosure even of unpublished patent applications, but only from a member who is also the proponent of a standard.

Curiously, virtually none of the organization rules I studied require a member to search either its own files or the broader literature to identify relevant intellectual property rights. Only three groups – the National Institute for Standards and Technology, the European Telecommunications Standards Institute, and the Frame Relay Forum – required such a search. ETSI's requirement is subject to waiver; it provides that either ETSI or the member will be required to conduct a search. Only the Frame Relay Forum specifies the sort of search that must be conducted, and even that is done in broad terms (a "reasonable" search). The failure of organizations to require searches, while understandable given the time and resource constraints under which members operate, gives rise to serious problems, as I discuss below.[24]

Most organizations permit members to own intellectual property rights in a standard, though they often discourage it. Only RosettaNet, which requires members to assign their intellectual property rights to the group, appears to flatly prohibit ownership of IP rights. Other groups discourage ownership of intellectual property without actually forbidding it, however. Thus, two groups studied (I2O SIG and Wired for Management) permit a member of own intellectual property, but only if they will license it to other members on a royalty-free basis. ISO requires members to give up patent rights, though not other sorts of IP rights. ETSI will reconsider its decision to approve a standard if the standard turns out to be controlled by an intellectual property right. The ATM Forum requires a ¾ majority to approve a standard governed by an intellectual property right, and similarly makes it easier to revoke a standard if it is found to be covered by intellectual property. Several organizations expressly discourage the ownership of intellectual property in standards, but will permit them in exceptional cases.[25] And at least one organization (the WAP Forum) appears to take inconsistent positions on the ownership of intellectual property.

Even those groups that permit members to own intellectual property rights covering a standard generally impose some conditions on the use of that intellectual property. The most common condition is that intellectual property rights be licensed on "reasonable and nondiscriminatory terms"; 14 of the 21 organizations with policies required members to license their patent rights on such terms. Another organization required outright assignment of patent rights, and two more required royalty-free licensing of patents.[26] Three organizations have a looser standard, requesting that members agree to license their patents on reasonable and nondiscriminatory terms, but not requiring that they do so.

While "reasonable and nondiscriminatory licensing" thus appears to be the majority rule among organizations with a patent policy, relatively few organizations give much explanation of what those terms mean or how licensing disputes will be resolved. Only two organizations specifically provide that the licensing obligation compels a member to license to everyone in the world using the standard, not just to license to other members. It seems unlikely that the remaining organizations intend to restrict the licensing obligation;[27] rather, it appears they simply haven't addressed the issue in their policies. Four organizations either give content to the obligation by specifying what a "reasonable" term means, or provide a mechanism for the organization to resolve disputes about license terms and fees. And one organization requires not only that a license term be reasonable and nondiscriminatory, but also that it not constitute "monopolistic abuse" of a patent. In short, while intellectual property owners at many organizations must license their rights on reasonable and nondiscriminatory terms, it is not clear what those obligations mean in practice.

---

[21] This was the policy of I20 SIG and Wired for Management. Similarly, RosettaNet required *assignment* of intellectual property rights to the organization itself, and the British Standards Institute relied on a provision of British patent law that gave licenses to members as of right. Only one group, the Distributed Management Task Force, did not require disclosure without some automatic provision for intellectual property owners giving up their rights.

[22] As we will see, this is a significant problem.

[23] In most of the world, patent applications are published 18 months after filing. In the United States, patent applications were until recently kept secret unless and until a patent issued. Beginning in 1999, most U.S. patent applications will be published 18 months after filing, though some patentees can maintain their application as a secret beyond that point. 35 U.S.C. §122.

[24] *See infra* __.

[25] *See* CEN/CENELEC (ownership of standards permissible "in exceptional cases" only); ANSI (patented standards acceptable only if "technical reasons justify this approach"); IETF ("prefers" unpatented technology).

[26] Two organizations, the Open Group and the ATM Forum, required royalty-free licensing of copyrights but permitted royalty-based licensing of patents. Thus, for copyrights the numbers in the text should read 12 of 21 policies requiring reasonable and nondiscriminatory licensing, and 4 of 21 requiring royalty-free licensing.

2. Implications of Diversity

The fact that different organizations have different rules governing intellectual property rights (or no rules at all) means that it is very difficult for intellectual property owners to know ex ante what rules will govern their rights. Because there is no standard set of rules, companies must investigate the by-laws of each organization they join in order to understand the implications of joining. While this doesn't seem that onerous a burden in the abstract, a number of practical considerations mean that companies are unlikely to be fully informed about their intellectual property position.

First, most companies in technology industries participate in more than one standard-setting organization. To take just one example, in 1998 Sun Microsystems participated in 87 different standards groups.[28]

**Figure 1.** *Standards Organization List by Function*

**General**
- Accredited Standards Committee (ASC) X3
- ANSI
- ECMA
- IEEE/POSIX
- International Organization for Standardization (ISO)
- International Telecommunication Union (ITU)
- ISO/IEC Joint Technical Committee 1 (JTC1)
- Open Group
  - Open Software Foundation (OSF)
  - X/Open
- X3

**OS and Architecture-Specific**
- ABI groups
- Large File Summit
- Power PC Group
- SPARC International
- Unicode Consortium
- UNIX International

**Communications and Networking**
- Asynchronous Transfer Mode (ATM) Forum
- CommerceNet
- Financial Services Technology Consortium (FSTC)
- Frame Relay Forum (FRF)
- Internet Society
  - Internet Engineering Tax Force (IETF)
  - Internet RFC's

**Communications and Networking, Cont.**
- Information Infrastructure Standards Panel (IISP)
- Network Management Forum (NMF)
- World Wide Web Consortium (W3C)

**Desktop and Graphics**
- Component Integration Laboratories (CIL)
- Desktop Management Task Force (DMTF)
- International Color Consortium (ICC)
- Interactive Multimedia Association (IMA)
- Moving Pictures Experts Group (MPEG)
- Multimedia and Hypermedia Information Coding Experts Group (MHEG)
- Open GL
- X Consortium

**US Government and Other Nations**
- Asia Oceanic Workshop (AOW)
- Chinese Open Systems Association (COSA)
- Defense Information Systems Agency (DISA)
- DoD Specifications and Standards
- European Commission DGXIII/E Open Information Interchange (OII) Initiative
- European Workshop for Open Systems (EWOS)
- Federal Information Processing Standards (FIPS)
- Japanese Industrial Standards Committee (JISC)
- National Institute of Standards & Technology (NIST)
- Open Systems Environment Implementors Workshop (OIW)

---

27 Such a restriction would pose serious antitrust concerns. *See infra* __.

28 *See* Figure 1. This list courtesy of Erv Basinski.

**Figure 1, cont.** *Standards Organization List by Function*

**Application**

- ANSA
- Electronic Data Interchange Association (EDIA)
- Electronic Messaging Association (EMA)
- Information Technology Association of America (ITAA)
- Open GIS Consortium (OGC)
- Open User Recommended Solutions (OURS)
- Petrotechnical Open Software Corporation (POSC)
- SQL Access Group – see X/Open
- UniForum Association

**Testing & Performance**

- Corporation for Open Systems (COS)
- Standard Performance Evaluation Corporation (SPEC)

**Documentation**

- Davenport Group
- Document Management Alliance (DMA)
- JTCI SC18/WG8
- SGML/Open

**Hardware**

- Electronic Industries Association (EIA)
- Open Firmware Working Group
- PCMCIA
- Plug and Play Association
- Rockridge Group
- Video and Electronics Standards Association (VESA)
- VMEbus International Trade Association (VITA)

**Objects**

- Object Database Management Group (ODMG)
- Object Management Group (OMG)

**Other Standards Groups**

- Association of Information and Image Management International (AIIM) – see DMA
- CAD Framework Initiative (CFI)
- Computer Systems Policy Project (CSPP)
- Data Management Interfaces Group (DMIG)
- Data Interchange Standards Association, Inc. (DISA)
- Digital Audio Visual Council (DAVIC)
- Document Enabled Networking (DEN) – see DMA
- International Multimedia Teleconferencing Consortium (IMTC)
- Message Oriented Middleware Association (MOMA)
- National Information Standards Organization (NISO)
- PEX
- QIC [tape cartridges]
- Shamrock Group – see DMA
- SPAG
- Telecommunications Information Networking Architecture Consortium (TINA-C)
- Transaction Processing Performance Council (TPC)
- Uniform Device Driver Interface Group (formerly CDDI)
- Workflow Mangement Coalition
- X.400 API Association (XAPIA)
- Z39.50 Information Retrieval Service and Protocol Standard

Similarly, there are dozens of different groups associated with Internet technical standards alone.[29]

[29] *See* Figure 2.

**Figure 2.** *Roadmap to the Communities and Parties of the Internet.*[30]

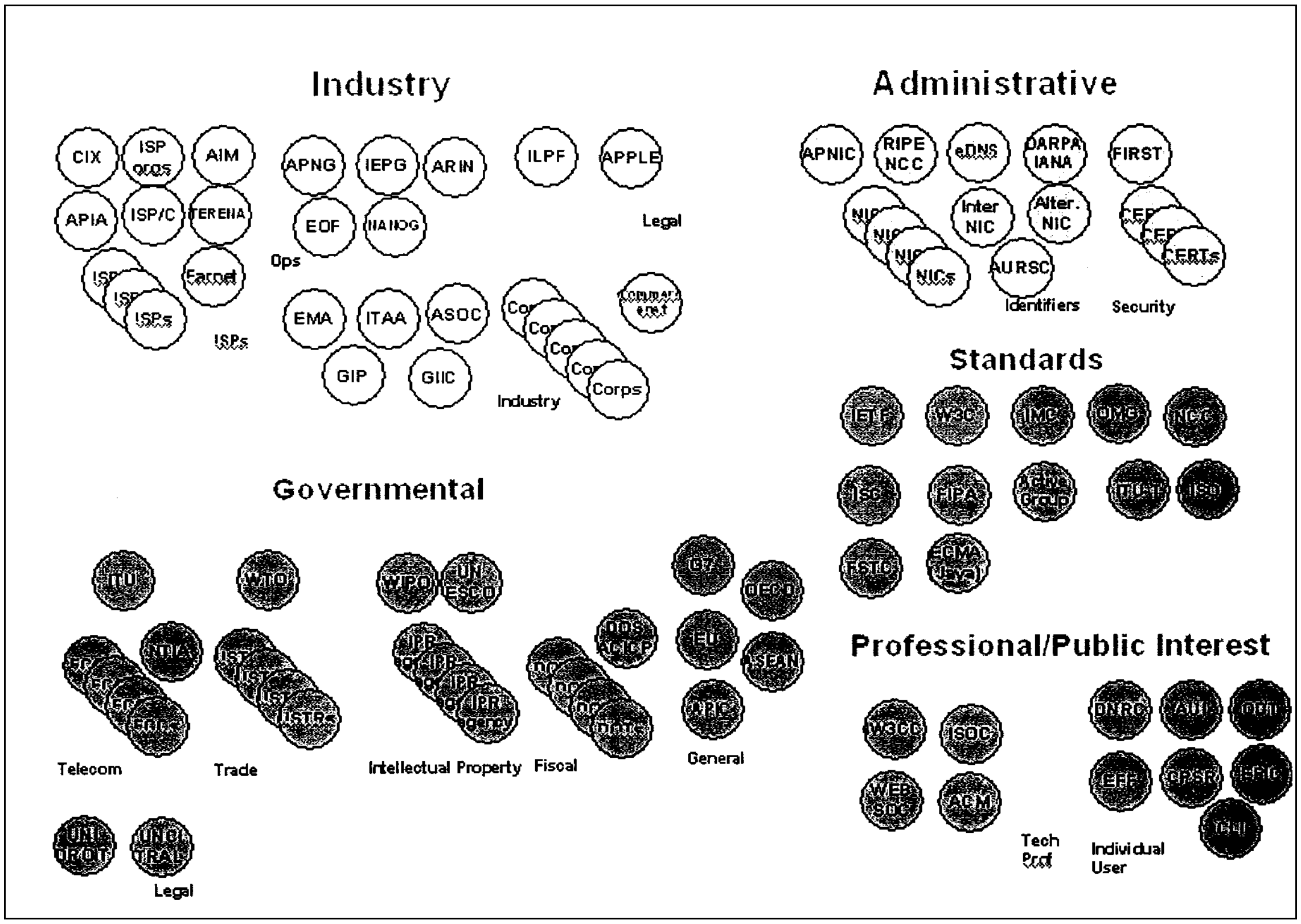

Thus, technology companies don't merely have to figure out what rules apply to them, but they face a labyrinth of different groups with overlapping subject matter concerns, each with its own set of rules. Because standard-setting organizations are concerned only with intellectual property rights that affect their particular standards, the likely result will be that some of a company's intellectual property rights will be subject to effective forfeiture, more will be subject to disclosure and licensing requirements, and some will not be restricted. Lawyers would have to examine each group, each standard, and each patent carefully to know for sure which is which.

This brings us to the second practical problem. Lawyers are rarely the ones to participate in standard-setting meetings. A company's representative to such a group is likely to be an engineer with little or no understanding of patent law. Indeed, in many cases the decision whether or not to join an organization is made at a fairly low level within a company, without the involvement of senior businesspeople, much less lawyers. If the organization in question is one of the few that compels assignment or royalty-free licensing, or requires a search for intellectual property, the decision to join may inadvertently commit the company to give up major intellectual property rights.

Because of these practical limitations, most technology companies today face a hodgepodge of rules and obligations of which they are only dimly aware. In the sections that follow, I explore some of the legal rules that bear on standard-setting organizations, identify some of the problems that arise in articulating and enforcing intellectual property policies, and offer some suggestions to all the players involved (members, organizations, courts, and scholars) for how to think about the intellectual property rules of standard-setting organizations.

30 Taken from http://www.wia.org/roadmap.htm.

**Table 1.**[30] *SSO Intellectual Property Rules*

| SSO | Policy? | Disclosure? | Search? | Can Standard Include IP? | Licensing Provisions |
|---|---|---|---|---|---|
| W3C[31] | P,TM,© | Yes | No | Yes | RAND requested but not required |
| I2O SIG[32] | P, TM | No | No | Yes[33] | royalty-free license required |
| Wired for Management[34] | P | No | No | Yes[35] | royalty-free license required for necessary claims only |
| IETF[36] | P,© | Yes | No | Yes[37] | RAND to all users; terms must be specified |
| IEEE[38] | P | Yes | No | Yes | RAND; terms must be specified |
| RosettaNet | P,© | No | No | No[39] | patents assigned to RosettaNet |
| IMC[40] | None | | | | |
| OMG[41] | all IP | Yes | No | Yes | RAND |
| ISC[42] | None | | | | |
| ITU[43] | P | Yes; includes pending patents | No | Yes | RAND; no "monpolistic abuse" |
| ISO[44] | P,TM,© | Yes | No | TM and © yes; patent no | patents must be given up or RAND required; nonexclusive copyright license to ISO; no trademark rule |
| FSTC[45] | None | | | | |
| NIST[46] | P | Yes | Yes | Yes | incorporates ANSI rules |
| ANSI[47] | P | Yes | No | Maybe[48] | RAND; ANSI will review claims of unreasonableness |
| ETSI[49] | P, utility model, designs | Yes | Depends[50] | Maybe[51] | RAND; irrevocable; but standard may be adopted even if patentee refuses to license |
| BSI[52] | P | No | No | Yes | users licensed as of right; British patent office to settle disputes as to terms[53] |
| ATM Forum | P,TM,© | Yes; includes only published applications | No | Yes[54] | royalty-free license as to copyrights; RAND licensing of patents or a written refusal to do so |
| CEN/CENELEC | P | Yes | No | Maybe[55] | RAND to entire world required or standard is withdrawn |
| Parlay Group | None | | | | |
| OGC | None | | | | |
| WAP Forum | P,© | Yes; includes some pending apps[56] | No | Unclear[57] | RAND required; possible public domain dedication |

**Table 1, cont.**[58] *SSO Intellectual Property Rules*

| SSO | Policy? | Disclosure? | Search? | Can Standard Include IP? | Licensing Provisions |
|---|---|---|---|---|---|
| DMTF[59] | P | No | No | Yes | RAND required or standard is withdrawn |
| MWIF[60] | all IP | Yes | No | Yes | royalty free license or RAND automatically compelled |
| OSGi[61] | all IP | Yes, including pending claims | No | Yes | RAND required by agreement to join group |
| Open Group[62] | P,© | Yes | De facto[63] | Yes | © must be licensed royalty-free; RAND for patents |
| CommerceNet | None | | | | |
| Frame Relay Forum | P | Yes; standards may be revoked for non-disclosure | "Reason-able" search required | Yes | RAND |
| AMI2[64] | None | | | | |
| JEITA[65] | In progress[66] | | | | |

30 In this table, P means patent, TM means trademark, © means copyright, RAND means "reasonable and nondiscriminatory licensing"
31 World Wide Web Consortium. *See* http://www.w3.org/Consortium/Legal/
32 Intelligent Input/Output Specification
33 Subject to the royalty-free license.
34 http://developer.intel.com/ial/wfm/wfmspecs.htm
35 Subject to the royalty-free license.
36 Internet Engineering Task Force. *See* RFC 1958, § 5.1; http://www.ietf.org/ipr.html. Some groups (such as the Organization for the Advancement of Structured Information Standards) adopt the IETF standards.
37 *See id.* ("prefer unpatented technology, but if the best technology is patented and is available to all at reasonable terms, then incorporation of patented technology is acceptable.").
38 Institute of Electrical and Electronics Engineers.
39 All intellectual property rights covered in a RosettaNet standard become the property of RosettaNet. Cargill email § 13.
40 Internet Mail Consortium
41 Object Management Group; http://www.omg.org
42 Internet Software Consortium, http://www.isc.org
43 International Telecommunications Union; http://www.itu.int/ITU-T/patent/Readme.html
44 International Organization for Standardization; http://www.iso.ch. For a detailed discussion of the ISO standard-setting process, see Mark A. Lemley & David McGowan, *Could Java Change Everything? The Competitive Propriety of a Proprietary Standard*, 43 **Antitrust Bull.** 715 (1998). A number of smaller groups (such as the UNICODE Consortium) explicitly adopt ISO/IEC rules.
45 Financial Services Technology Consortium; http://www.fstc.org
46 National Institute of Standards and Technology; http://www.nist.gov
47 American National Standards Institute; for a discussion of the ANSI policy, see Robert P. Feldman & Maura L. Rees, *The Effect of Industry Standard Setting on Patent Licensing and Enforcement*, **IEEE Communications**, July 2000, at 112, 113. A number of smaller technical groups (VITA Standards Organization covering the VME Bus standard) explicitly adopt the ANSI approach.
48 ANSI permits patented standards only if "technical reasons justify this approach." ANSI, *Procedures for the Development and Coordination of American National Standards*, §1.2.11.1.
49 European Telecommunications Standards Institute. For a discussion of ETSI policy, see Johan Verbruggen & Anna

Lorincz, *Patents and Technical Standards* §3.1.B (working paper 2001).

[50] EC policy requires that the patent owner conduct a search unless the standard-setting body commits to do the search itself. *See* European Commission, *Communication on IPRs*, ¶¶4.5.1, 4.5.2.

[51] ETSI policy provides that the General Assembly shall refer cases of patent ownership to the EC and EFTA "for their consideration" if the patentee refuses to license on reasonable and nondiscriminatory terms. *Id* at 15.

[52] British Standards Institute. *See* Verbruggen & Lorincz, *supra* note __, at 16.

[53] This is pursuant to the U.K. Patent Act of 1977, § 46.

[54] If the patentee has refused to grant a license to patents covering a proposed standard on reasonable and nondiscriminatory terms, a ¾ vote of the membership is required to approve the standard. If the standard has already been issued when the problem arises, 1/3 of the members may vote to revoke the standard. ATM Forum Bylaws, Article 3.12.2.d.

[55] A standard may include patented technology "in exceptional cases" only. CEN/CENELEC Memorandum No. 8, *Standardization and Intellectual Property Rights* § 1 (1992).

[56] Proponents of a standard must notify WAP when an application is filed; other members need not do so, but if intellectual property is put into a standard it is "in public domain" and should not be subsequently patented. Cargill email ¶5.

[57] On the one hand the WAP Forum standard says intellectual property included in a standard that is accepted is "in the public domain," but on the other hand it also speaks of licensing on reasonable and nondiscriminatory terms. *Id*.

[58] In this table, P means patent, TM means trademark, © means copyright, RAND means "reasonable and nondiscriminatory licensing"

[59] Distributed Management Task Force. *See* Cargill email ¶6.

[60] Mobile Wireless Internet Forum. *See* Cargill email ¶14.

[61] Open Services Gateway Initiative. *See* Cargill email ¶16.

[62] http://archive.opengroup.org/itdialtone/architecture/procedures/iac.htm.

[63] While the Open Group rules do not require a search, they do require patentees to agree not to sue users of the standard for any patents that were not disclosed during the process. This has an effect analogous to a search requirement.

[64] Advanced Memory Int'l, Inc., http://www.ami2.org/

[65] Japan Electronics and Information Technology Industries Association. http://www.eiaj.or.jp/english/index.htm.

[66] Site viewed Jan. 9, 2001.